\documentstyle[11pt,paspconf]{article}

\begin{document}

\title{Rings and Echoes: An Overview$^1$}

\author{Arlin P.S.~Crotts$^2$}
\affil{Department of Astronomy, Columbia University, New York, NY~~10027
U.S.A.}

\altaffiltext{1}{
Based in part on observations made with the NASA/ESA {\it Hubble Space
Telescope}, obtained from the data archive at the Space Telescope Science
Institute, which is operated by AURA, Inc., under NASA contract NAS 5-26555.}

\altaffiltext{2}{Visiting Astronomer, Cerro Tololo Inter-American Observatory. 
CTIO is operated by AURA, Inc., under cooperative agreement with the National
Science Foundation}

\begin{abstract}

The volume around the SN~1987A contains a variety of structures, including
but certainly not limited to the three-ring nebula glowing in recombination
lines.
Many of these structures are revealed by light echoes, hence have been mapped
in three dimensions by our optical imaging monitoring of the field around the
SN.
The three rings are part of a bipolar nebula which contains them at its waist
and crowns, and which is itself contained in a larger, diffuse nebula with a
detectable equatorial overdensity.
This diffuse nebula terminates in a boundary overdensity which most likely
marks the inner edge of a bubble blown by the main sequence wind of the
progenitor star and its neighbors.
Beyond this bubble is a rich collection of interstellar structures revealed by
light echoes.
In addition to detecting and mapping these structures, we add dynamical and age
information by establishing the kinematics of the gas, both on interstellar and
circumstellar scales.
These reveal, for instance, a timescale for the outer circumstellar rings which
is in close agreement with the inner ring.
The presence of these structures, their ages and morphologies should be
included in any model explaining the evolution of the progenitor star and its
mass loss envelope.
\end{abstract}

\keywords{}

\section{Introduction}

The nebula around SN 1987A is a complex structure, consisting not only of the
three rings glowing in recombination radiation now seen so easily by {\it HST},
but several additional layers of structure seen by other methods, as well as
further structure inferred by the behavior of the already interacting SN
ejecta.
Regarding the rings, the presence of material from the inner ring was
discovered as narrow emission lines by IUE in 1987 (Fransson et al.~1989), and
shown to be spatially resolved from the ground (Wampler \& Richichi 1989).
The outer rings (the northern ``NOR'' and southern ``SOR'') and the
approximate shape of the inner ring (``ER'' for equatorial ring) were
discovered by Crotts, Kunkel and McCarthy (1989).
The structure of the rings was further resolved by observations at the NTT
(Wampler et al.~1990) and {\it HST} (Jakobsen et al.~1991, Burrows et
al.~1995).
The kinematics of the ER was shown (Crotts \& Heathcote 1991) to indicate a
true ring, and not just a limb-brightened spheroid.
The three rings are connected by a double-lobed nebula containing the ER at its
central waist and probably terminating at its outer extremes at the positions
of the ORs (Crotts, Kunkel \& Heathcote 1995).

A light echo just outside the three rings was discovered by Bond et al.~(1991)
and shown by Crotts and Kunkel (1991) to encompass a diffuse medium of echoing
material.
Also incorporated into this diffuse medium is the discontinuity in density (or
at least in the slope of the radial density fall-off) known as ``Napoleon's
Hat'' (Wampler et al.~1990), and an equatorial dust overdensity in the same
plane as the ER (Crotts, Kunkel \& Heathcote 1995).

In this paper we review some of this work and present some previously
unpublished results that bear on the question of the progenitor star's nature
and the production of the circumstellar nebula.

\section{Kinematics of the Three Rings}

From the velocity field, measured in [N~II]$\lambda$6583 emission,
it has been established that the ER expansion produces a gradient across the
minor axis as would be caused by a ring, inclined at $43^\circ$, expanding
radially at $v_{exp} = 10.3$~km~s$^{-1}$ (Crotts \& Heathcote 1991).
Measurements made by other groups correspond to expansion velocities of 10.3,
8.3 and 11~km~s$^{-1}$ (Cumming 1994, Meaburn, Bryce \& Halloway 1995, and
Panagia et al.~1996, respectively).
Here we present similar data from a series of similar measurements obtained by
Heathcote \& Crotts at the CTIO 4-meter since March 1989, 10 of which have been
analyzed at present.
These yield radial expansion values in the range $9 < v_{exp} <
10.6$~km~s$^{-1}$.

\begin{figure}
\vspace{2.55in}
\caption{
a: (left panel) The unmodified echelle spectrum (excepting bias subtraction,
flat-fielding and summing) of an 18-arcsec $\times$ 4.7\AA\ region centered on
the [N~II]$\lambda$6583 emission line from SN~1987A at PA 130$^\circ$.
This was taken with the CTIO 4-meter/echelle by S.~Heathcote on day 3245.
~~~b: (right panel) An expanded view of the central 12 arcsec $\times$ 3.2\AA\,
once the signals from continuum sources and extended interstellar features
have been removed. 
Note the extensions from each of the four corners of the emission locus.
These are not present in the ER-only model for the nebula's spectrum.
} \label{fig-1}
\end{figure}

More novel, however, is the combination of angular resolution at some epochs
and spectral resolution which allows the ER and OR signals to be separated.
This includes several stages of analysis:
1) the continuum from the SN plus Stars 2 and 3 (at some position angles) must
be removed by interpolation in wavelength across the [N~II] emission line;~
2) interstellar [N~II] emission from several components, particularly
255, 277, 289 and 313~km~s$^{-1}$ must be modeled and also removed by
interpolation.
We note the presence of a component with $259$~km~s$^{-1} <v< 301$~km~s$^{-1}$
extending to about 20 arcsec from the SN, which we discuss elsewhere;~
3) the signal from the ER must be modeled in detail, accounting for the ring
geometry and flux non-uniformity, seeing, wavelength resolution, gas
temperature and spectrograph point-spread-function, leading to a fit which is
good to about 10\% in local surface brightness for the worst residuals in the
inner 99\% of the enclosed energy from the ER signal.
Beyond this region, once the ER model is subtracted, there are residual signals
that persist at $\ga$100\% of the local ER level.
We suspect that these are real because not only do they persist even if the
ER model is over-subtracted, but they change positions depending on the
position angle (PA) of the spectrograph slit relative to the sky.
(However, when the slit is rotated $180^\circ$, the residuals maintain the same
relative positions.)~
4) These residuals are then fit with a 2-D elliptical gaussian and the centroid
velocity is noted.

\begin{figure}
\vspace{2.45in}
\caption{Contours of residual signal once the model for the ER is subtracted
from the data shown in Figure 1b (a 9-arcsec $\times$ 2.3\AA\ subregion).
I have intentionally presented a case where the ER is slightly oversubtracted;
nonetheless, the positive residuals beyond the ER remain.
For comparison, the highest counts before subtraction were about 2500 ADU.
} \label{fig-2}
\end{figure}

When one plots the positions and velocities of these outer positive residuals,
one sees that their positions fall near the
ORs, and that the values observed are consistent with neighboring values.
Also, if one assumes that the ORs sit near the crowns of the bipolar nebula
from Crotts, Kunkel \& Heathcote (1995), and are expanding radially from the
SN, one would expect the northern and southernmost extents of the ORs to be
seen at nearly zero velocity relative to the SN.
The southern edge of the NOR should be expanding most rapidly away from Earth,
while the northernmost SOR portions should be moving most quickly towards
Earth.
This is seen, in fact, in Figure 3.

For the NOR, we find a range of velocities, relative to the 289~km~s$^{-1}$
centroid of $+1$ to $+26$~km~s$^{-1}$, and for the SOR 0 to $-26$~km~s$^{-1}$.
Using the minor and major axis lengths measured from the day 2770 WFPC2
image in the F658N filter (primarily [N~II]$\lambda 6583$, but with significant
contamination from H$\alpha$), and assuming these rings are round, their axis
ratios correspond to $44^\circ$ for the NOR and $31^\circ$ for the SOR.
Note that the SOR is demonstrably non-elliptical, being closer to circular on
the western versus the eastern side.
We also consider the possibility that the SOR is distorted, and lies in a plane
inclined at the same angle as the ER ($43^\circ$), nearly the same inclination
as the NOR.
The major axes of the NOR and SOR correspond to distances of $d_{maj}=
0.83$~pc and $0.86$~pc, respectively, assuming a
distance of 50~kpc to SN~1987A.

\begin{figure}
\vspace{2.6in}
\caption{Outline of ER and ORs, with PA of three slit positions shown.
(PA 20 is commonly observed instead of PA 10.)~
Plotted against these are the centroid positions and velocity of positive flux
residuals like those seen in Figure 2.
ER velocities are {\it not} shown.
The PA 60, SOR measurement of $-1$~km~s$^{-1}$ is uncertain.
} \label{fig-3}
\end{figure}

Given a velocity difference $\Delta v$ of 25~km~s$^{-1}$ for the NOR and
26~km~s$^{-1}$ for the SOR, and the inclinations above, one finds an expansion
velocity $v_{rel} = 36.0$~km~s$^{-1}$ for the NOR, and 38.1~km~s$^{-1}$ and
50.5~km~s$^{-1}$ for the SOR, for $i=43^\circ$ or $31^\circ$, respectively.
The ratio $d_{maj}/v_{rel}$ is a measure of the age of these structures,
assuming uniform expansion, which is 22500~y for the NOR, and 22000~y or
16600~y for the SOR, for $i=43^\circ$ or $31^\circ$, respectively.
(Note that these numbers are somewhat larger than those presented at the
workshop; in the intervening time I have remeasured some quantities and
corrected a simple error.)~
A similar measurement for the ER corresponds to 20000~y (Crotts \& Heathcote
1991) and is consistent with values for the NOR, and for the SOR at
$i=43^\circ$.
The near-equality of the $d_{maj}/v_{rel}$ timescale for the ER and ORs
corresponds to the expectation from homologous expansion.

The full implications of these results await complete numerical modeling
describing the evolution of the progenitor until the explosion,
however, a simple interpretation is still useful.
These velocities are somewhat at odds with spectroscopy of the NOR from the
{\it HST} FOS (Panagia et al.~1996), which indicate that the ORs are depleted
in nitrogen compared to the ER.~
Panagia et al.~interpret this underabundance as the result of the OR material
being ejected about $10^4$y before the ER.
This is not supported by a simple interpretation of the kinematics of the
rings.

The expansion of the NOR at 26~km~s$^{-1}$ along the line of sight over 22000~y
corresponds to an extent of 0.59~pc behind the SN.
The most distant portion of the bipolar nebula mapped by echoes sits at nearly
the same point, 0.65~pc behind the SN.
(This match is also obtained in AAT echelle data on the rings: Cumming \&
Lundqvist [1996].)~
This supports the idea that the ORs formed at the outer boundaries of the
bipolar nebula, which has been used as evidence for the presence of an H~II
region interior to the bipolar nebula, leading to the early X and radio rise in
the luminosity of the SN (Chevalier \& Dwarkadas 1995).

\section{Echoing Circumstellar Matter Beyond the Rings}

Image subtraction techniques which produced the three-dimensional maps of
circumstellar (Crotts, Kunkel \& Heathcote 1995) and interstellar structure
(Xu \& Crotts \& Kunkel 1995) in the vicinity of SN~1987A have been applied
to the region just outside the bipolar/triple-ring nebula, revealing the
evolution of four features:
1) a ring at radii of 9-15 arcsec (depending on the epoch of observation),
discovered by Bond et al.~and modeled by Chevalier and Emmering (1989) as a
contact discontinuity between the red supergiant (RSG) wind and the surrounding
bubble blown by the main-sequence (MS) progenitor wind.
For the sake of discussion, we will use a more general term ``circumstellar
boundary'' (CB);~
2) a sheet of material laying along the equatorial plane defined by the ER and
bisecting the bipolar nebula (c.f.~Crotts, Kunkel \& Heathcote 1995,
Fig.~21);~
3) a diffuse echo (Crotts \& Kunkel 1991) extending from the bipolar nebula
outward to feature \#1, and~
4) ``Napoleon's Hat,'' (Wampler et al.~1990) apparently a
discontinuity in the gradient of feature (3).
We have followed these features since 1988 or 1989 until their disappearance
at various times since.

\begin{figure}
\vspace{2.7in}
\caption{
The residual image showing, in order of increasing radius, echoes from the
diffuse circumstellar medium, circumstellar boundary and ISM 130~pc away from
the SN.
With permanent nebulosity and the net flux from stars removed (but some stellar
residuals remaining - including from Star 2), the echoes become more visible.
Contained within the diffuse echo is the bipolar/triple-ring nebula.
This image was obtained on day 750 at the LCO 2.5-meter by W.~Kunkel, in a
band centered at 6023\AA.
The field of view is 85 arcsec in width.
} \label{fig-4}
\end{figure}

While the echo from the CB was flocculent in appearance, its individual
components tended to lie in well-defined ellipses, until the feature began
fading drastically in 1991-2.
Beyond these epochs we find the locus of the overall echo difficult to trace.
(We are trying to to improve the signal-to-noise of the residual image by using
PSF-matching techniques e.g.~Crotts \& Tomaney [1996], but cannot report these
results yet.)~
In the meantime we simply centroid on individual echo patches, and display
them in three dimensions ($x$ and $y$ on the plane of the sky, and
line-of-sight depth $z=(x^2+y^2)/2ct-ct/2$, where $t$ is time since maximum
light).

Figure 5 shows two views of the CB (as well as the bipolar nebula near the
origin) as seen from vantage points perpendicular to the sightline to Earth.
Figure 5a shows the view from far to the north.
The CB wraps around the SN at a radius of about 5~pc.
The only points sampled must be contained within the light-echo parabolae
for the epochs of observation.
Points close to the SN in $x$ can appear inside the paraboloids.
Note that in later epochs, from 1992-3, the CB appears to be fading away, as
evidenced by the lack of points at the location of the largest parabolae.

In Figure 5b we see the view of the CB (and bipolar echo) from the vantage of
an observer far to the east.
It is evident that the CB viewed in this direction shows a larger radius of
curvature than when viewed far to the north, as in Figure 5a.
Possibly the CB is elongated, in the sense that the region
radially outward from the southern pole of the bipolar nebula, perpendicular to
the ER plane, is at a larger distance from the SN than points near the ER
plane.
The dashed line extending from the northern end of the bipolar nebula to the CB
at the same $y$ as the SN shows the position of Napoleon's Hat, derived
consistently from data at several epochs.
Note that it does NOT extend beyond the CB.

The fading of the CB (as well as the diffuse echo) can be understood in terms
of the behavior of a Henyey-Greenstein (1941) model dust with phase factor
$g = \overline {cos ~ \alpha}$ and the related scattering function
$F(\alpha) = {1-g^{2}} / (1+g^{2}-2g ~ cos ~ \alpha)^{3/2}$.
For $0.6 < g < 0.8$, the observed fading of the echoes with
increasing scattering angle $\alpha$ should be expected, and these are
reasonable $g$ values for interstellar dust, at least in the Galaxy
(c.f.~Hurwitz, Bowyer \& Martin 1991, Witt 1989).

The implications of these structures are significant.
The fact that the Napoleon's Hat feature fails to extend beyond the CB argues
against it composing a portion of a bow shock (Wang, Dyson \& Kahn 1993) due to
the interaction of the SN moving through the interstellar medium (ISM): the
northern half of the CB sits relatively undisturbed beyond Napoleon's Hat,
continuing the structure established in the southern half of the CB.
We should note that a diffuse echo arises from either side of Napoleon's Hat
(denser on the interior); indeed, the feature appears to be a change in slope
of the radial fall-off in dust density.
Perhaps this is due to a change in outflow velocity, or a minor shock feature
due to slower, low density RSG material being overtaken by faster RSG wind
phase.
It is at least an indication of anisotropic, time-variable outflow.

\begin{figure}
\vspace{3.0in}
\caption{The 3-D geometry of the circumstellar boundary relative
to the bipolar nebula and the volume probed by light echoes.~~~
a: (left panel) the view from far to the north ($x$ is distance in right
ascension).
The CB wraps around the SN at a radius of about 5~pc.
The points sampled must be contained within the light-echo parabolae
for the epochs of observation.
~~~b: (right panel) view of the CB and bipolar echo for
an observer far to the east ($y$ is distance in declination).
The thick dashed line extending from the northern end of the bipolar nebula
to the CB at the same $y$ as the SN shows the position of Napoleon's Hat.
The nearly diagonal, thin dashed line shows the approximate symmetry axis of
the inner nebula.
} \label{fig-5}
\end{figure}

If the CB is actually a contact discontinuity (Chevalier \& Emmering 1989)
between the RSG wind and the ISM bubble blown by the MS blue supergiant (BSG)
progenitor, the pressure in this bubble should be nearly uniform
over a few parsecs, and the shape of the CB is due primarily to the
ram pressure $\rho v^2$ against the nearly uniform pressure $P$ of the bubble.
If the CB is not spherical, this is due to $\rho v^2$ being anisotropic.
Given the equatorial overdensity (feature \# 2 mentioned above) in the RSG
wind, a polar bulge or even a spherical CB indicates that the velocity of the
RSG wind at the poles must be higher than at the equator.
This is a wind configuration that has not yet been considered in interacting
wind models for the SN progenitor, and might lead to qualitatively different
results than those published e.g.~Blondin \& Lundqvist (1993), Martin \& Arnett
(1995), Blondin, Lundqvist \& Chevalier (1996).

The presence of the CB and the diffuse material inside it runs counter to some,
more complicated geometries proposed for the circumstellar environment of
SN~1987A (c.f.~Podsiadlowski, Fabian \& Stevens 1991); we see a continuously
distributed, higher density region (at least in terms of its echo reflectivity)
bounded on its interior and exterior by low density regions.
While this corresponds to the approximate morphology of a BSG - RSG - BSG
interacting wind model, the shape of the inner bipolar nebula and
rings, and now the outer CB, have yet to be successfully modeled.
Still, the presence and shape of this diffuse medium presents a strong
challenge to models not dominated by the basic BSG - RSG - BSG evolutionary
picture.

This diffuse medium gives some clue as to the timescale of the RSG mass loss
phase.
If the velocities seen at the inner edge of this wind e.g.~at the ER, NOR and
SOR apply throughout, $v_{exp} \approx 20$~km~s$^{-1}$, and if the RSG proceeds
unshocked to the CB (at least on the side unaffected by Napoleon's Hat), the
age of this diffuse structure is then $2\times 10^5$y.
The mass of dust in the CB is about twice that in the diffuse nebula (Crotts \&
Kunkel 1991), however, so the age of the overall nebula and of the RSG mass
loss phase may be several times greater, depending on how mass loss ${\dot M}$
evolves.

\section{Other Developments}

Several topics presented at the workshop which cannot be treated in this space
include our analysis of the propagation of the ultraviolet pulse through the
bipolar nebula, the measurement of the reflected echo from the UV pulse, and
the dynamics and kinematics of the interstellar medium in front of SN~1987A,
studied by combining echo 3-D mapping with velocity maps created from long-slit
echelle spectroscopy.
These studies can be found in Plait and Crotts (1997), Crotts et al.~(1997a, b)
and Xu and Crotts (1997), respectively.

At the workshop, I also reported briefly on our interstellar echo work,
detailed in Xu, Crotts and Kunkel (1996) and references therein.
A primary result is our placement using echoes of a large amount of material
at large distances, up to 1~kpc, in front of SN~1987A.
Of particular importance to several other investigations is the placement of
the SN approximately 500~pc behind the massive OB association LH~90, which
defines the centroid mass in this region of the LMC, presumably.
This implies that the SN sits well behind the LMC disk, presumably in a low
gas density environment.
We speculate, since our age for LH~90 and surrounding superbubble N157C are
nearly identical to that of the SN progenitor, that all three were created
together.
In which case, the SN has decelerated somewhat to its redshift of
20~km~s$^{-1}$ relative to N157C in order to reach its current position in
7~My.

\acknowledgements
This presentation reflects the major efforts of my collaborators, in this case
Bill Kunkel, Steve Heathcote and Jun Xu.
We are grateful for the continued support of Cerro Tololo and Las Campanas
observatories for our observing campaign.
We would also like to express our appreciation for the efforts of the workshop
organizers.


\begin{references}
\reference Blondin, J.M.~\& Lundqvist, P.~1993, \apj, 405, 337
\reference Blondin, J.M., Lundqvist, P.~\& Chevalier, R.A.~1996, \apj, 472, 257
\reference Bond, H.E., Gilmozzi, R., Meakes, M.G.~\& Panagia, N.~1990, \apjlett, 354, L49
\reference Burrows, C., et al.~1995, \apj, 452, 680
\reference Chevalier, R.A.~\& Dwarkadas, V.V.~1995, \apjlett, 452, L45
\reference Chevalier, R.A.~\& Emmering, R.T.~1989, \apjlett, 342, L75
\reference Crotts, A.P.S., Ghosh, H., Landsman, W.B, Stecher, T.P.~\& the UIT Team 1997a, in preparation
\reference Crotts, A.P.S.~\& Gilmozzi, R.~1997b, in preparation
\reference Crotts, A.P.S.~\& Kunkel, W.E.~1991, \apjlett, 366, L73
\reference Crotts, A.P.S., Kunkel, W.E.~\& Heathcote, S.R.~1995, \apj, 438, 724
\reference Crotts, A.P.S., Kunkel, W.E.~\& McCarthy, P.J.~1989, \apjlett, 347, L61
\reference Crotts, A.P.S.~\& Heathcote, S.R.~1991, Nature, 350, 683
\reference Crotts, A.P.S.~\& Tomaney, A.B.~1996, \apjlett, 473, L87
\reference Cumming, R.J.~1994, Ph.D.~thesis (Imperial College)
\reference Cumming, R.J.~\& Lundqvist, P.~1996, in Advances in Stellar
Evolution, R.T.~Rood (Cambridge University Press), in press
\reference Fransson, C., et al.~1989, \apj, 336, 429
\reference Henyey, L.G.~\& Greenstein, J.L.~1941, \apj, 93, 70
\reference Hurwitz, M., Bowyer, S.~\& Martin, C.~1991, \apj, 372, 167
\reference Jakobsen, P., et al.~1991, \apjlett, 369, L63
\reference Martin, C.~\& Arnett, D.~1995, \apj, 447, 378
\reference Meaburn, J., Bryce, M.~\& Holloway, A.J.~1995, \astap, 299, L1
\reference Panagia, N., et al.~1996, \apj, 459, 17
\reference Plait, P.~\& Crotts, A.P.S.~1997, \apj, submitted
\reference Podsiadlowski, Ph., Fabian, A.C.~\& Stevens, I.R.~1991, Nature, 354, 43
\reference Wampler, E.J.~\& Richichi, A.~1989, \astap, 217, 31
\reference Wampler, E.J., et al.~1990, \apjlett, 362, L13
\reference Wang, L., Dyson, J.E.~\& Kahn, F.D.~1993, \mnras, 261, 391
\reference Witt, A.N.~1989 in Interstellar Dust, L.~Allamandola \& A.~Tielens
(Dordrecht: Kluwer), p. 87
\reference Xu, J., Crotts, A.P.S.~\& Kunkel, W.E.~1995, \apj, 451, 806 (Erratum: 463, 391)
\reference Xu, J.~\& Crotts, A.P.S.~1997, \apj, in press
\end{references}
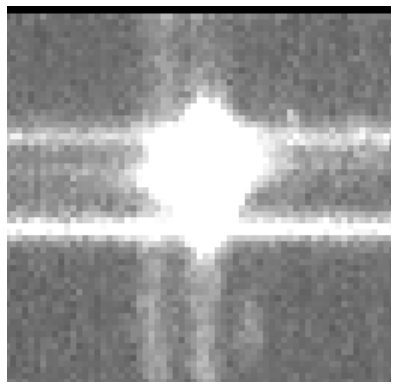
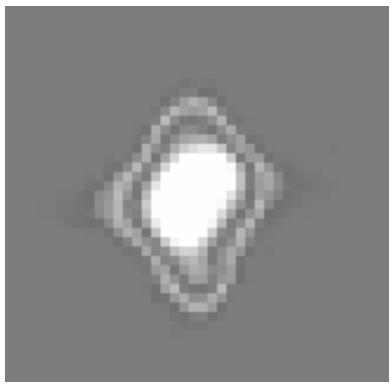
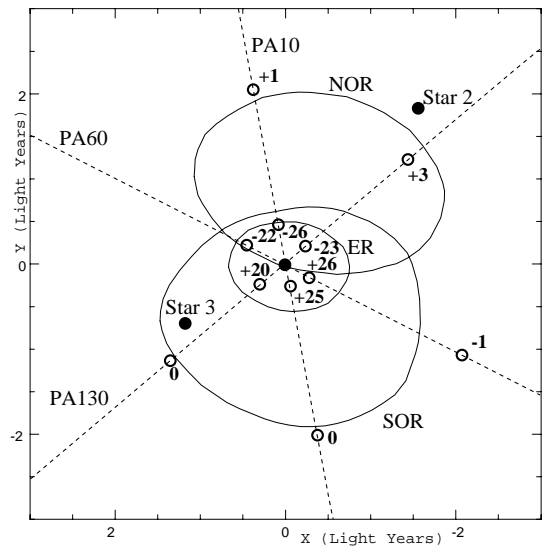
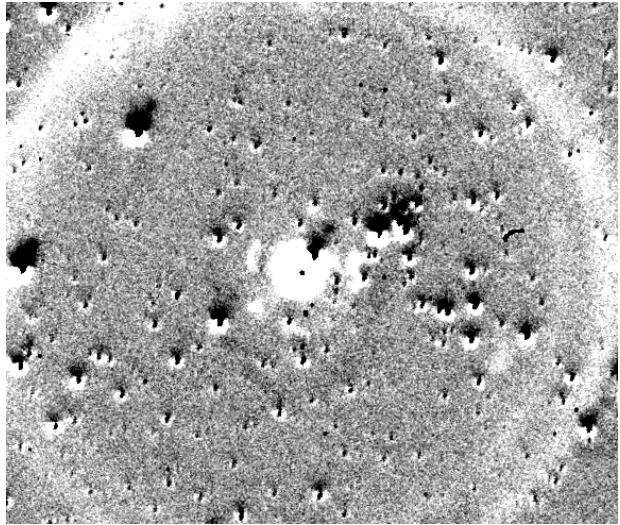
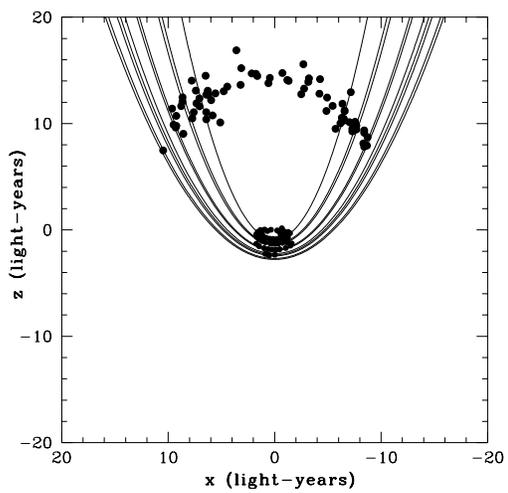
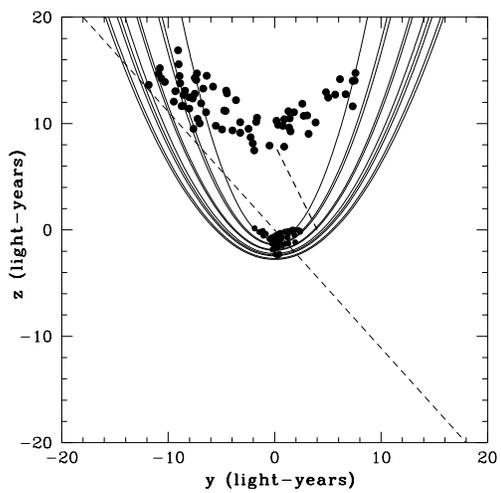
\end{document}